\documentclass[5p,longtitle,review]{elsarticle}

\usepackage{amsmath}
\usepackage{txfonts}
\usepackage{graphicx}
\usepackage{comment}
\usepackage{natbib}
\bibpunct{(}{)}{;}{a}{}{,}

\newcommand{\pmi}{$\pm$}

\begin{document}
\begin{frontmatter}

   \title{Measurement of the Crab Nebula spectrum over three decades
     in energy with the MAGIC telescopes}


%
\author[1]{J.~Aleksi\'c}
\author[2]{S.~Ansoldi}
\author[3]{L.~A.~Antonelli}
\author[4]{P.~Antoranz}
\author[5]{A.~Babic}
\author[6]{P.~Bangale}
\author[7]{J.~A.~Barrio}
\author[8,27]{J.~Becerra Gonz\'alez}
\author[9]{W.~Bednarek}
\author[10]{E.~Bernardini}
% ...
\author{(The MAGIC collaboration)}
\author[20]{D.~Horns}
\author[14]{J.~{Mart{\'{\i}}n}}
\author[26]{M.~Meyer}

\address[1]{IFAE, Campus UAB, E-08193 Bellaterra, Spain}
\address[2]{Universit\`a di Udine, and INFN Trieste, I-33100 Udine, Italy}
\address[3]{INAF National Institute for Astrophysics, I-00136 Rome, Italy}
\address[4]{Universit\`a  di Siena, and INFN Pisa, I-53100 Siena, Italy}
\address[5]{Croatian MAGIC Consortium, Rudjer Boskovic Institute, University of Rijeka and University of Split, HR-10000 Zagreb, Croatia}
\address[6]{Max-Planck-Institut f\"ur Physik, D-80805 M\"unchen, Germany}
\address[7]{Universidad Complutense, E-28040 Madrid, Spain}
\address[8]{Inst. de Astrof\'isica de Canarias, E-38200 La Laguna, Tenerife, Spain}
\address[9]{University of \L\'od\'z, PL-90236 Lodz, Poland}
\address[10]{Deutsches Elektronen-Synchrotron (DESY), D-15738 Zeuthen, Germany}
% ...
\address[14]{Institute of Space Sciences, E-08193 Barcelona, Spain}
% ...
\address[20]{Institut f\"ur Experimentalphysik Univ. Hamburg, D-22761 Hamburg, Germany}
% ...
\address[26]{Stockholm University, Oskar Klein Centre for Cosmoparticle Physics, SE-106 91 Stockholm, Sweden}
\address[27]{now at: NASA Goddard Space Flight Center, Greenbelt, MD 20771, USA and Department of Physics and Department of Astronomy, University of Maryland, College Park, MD 20742, USA}
% ...

\date{Accepted version 29.0, January 29, 2015}


\begin{abstract}
The MAGIC stereoscopic system collected 69 hours of Crab Nebula
data between October 2009 and April 2011. Analysis of this data
sample using the latest improvements in the MAGIC stereoscopic
software provided an unprecedented precision of spectral and
night-by-night light curve determination at gamma rays.
We derived a differential spectrum with a single instrument from 50 GeV up
to almost 30 TeV with 5 bins per energy decade. At low energies, 
MAGIC results, combined with \emph{Fermi}-LAT data, show a flat and broad
Inverse Compton peak. The overall fit to the data between 1\,GeV and 30\,TeV 
is not well described by a log-parabola function.
We find that a modified log-parabola function with an exponent of 2.5 instead
of 2 provides a good description of the data ($\chi^2_\mathrm{red} = 35/26$).
Using systematic uncertainties of the MAGIC and \emph{Fermi}-LAT measurements
we determine the position of the Inverse Compton peak to be at ($53 \pm
3_{\mathrm{stat}} +31_{\mathrm{syst}} -13_{\mathrm{syst}}$)\,GeV, which is the
most precise estimation up to date and is dominated by the systematic effects.
There is no hint of the integral flux variability on daily scales at
energies above 300 GeV when systematic uncertainties are included in the flux
measurement.
We consider three state-of-the-art theoretical models to describe the overall
spectral energy distribution of the Crab Nebula.  The constant B-field
model cannot satisfactorily reproduce the VHE spectral measurements presented
in this work, having particular difficulty reproducing the broadness of the
observed IC peak. Most probably this implies that the assumption of the
homogeneity of the magnetic field inside the nebula is incorrect.
On the other hand, the time-dependent 1D spectral model provides a good fit of
the new VHE results when considering a 80\,$\mu$G magnetic field. However, it
fails to match the data when including the morphology of the nebula at lower
wavelengths.

\end{abstract}


\end{frontmatter}

\section{Introduction}
The Crab pulsar wind nebula (PWN) is a leftover of the supernova
explosion that occurred in 1054 A.D. \citep{stephenson:2003a}, and it
is powered by the pulsar PSR B0531+21 at its center
\citep[for a detailed review]{hester:2008:aara}.
The Crab Nebula continuously supplies relativistic particles, mainly positrons and
electrons, that advect in the magnetized wind of the neutron star.
These relativistic particles are thought to be accelerated to a
power-law distribution either via a Fermi-like acceleration process
taking place at the termination shock (TS) \citep[and references therein]{arons:1994} or via shock-driven reconnection
in a striped wind \citep{petri:2007,sironi:2011:stripedwind}.
The downstream flow interacts with the
surrounding magnetic and photon fields creating the PWN. The nebula
emits synchrotron radiation which is observed from radio frequencies
up to soft $\gamma$ rays. This emission is well described by the
magnetohydrodynamic (MHD) model of \citet{kennel:1984a:crab}. At higher
energies (above 1 GeV), the overall emission is instead dominated by the Inverse
Compton (IC) up-scattering of synchrotron photons by the relativistic
electrons in the nebula \citep{deJager:1992:crab,atoyan:1996a:crab}. 

The Crab Nebula is one of the best studied objects in the sky.
Due to its brightness at all wavelengths, precise measurements
are provided by different kinds of instruments, allowing for many
discoveries, later seen in other non-thermal sources, and a 
detailed examination of its physics \citep[for a detailed review]{buehler:2014:review}.
The IC emission from the Crab Nebula was detected for the first time
above 700 GeV by the pioneering Whipple imaging atmospheric Cherenkov telescope in 1989
\citep{weekes:1989a}. Since then, the imaging Cherenkov technique 
  has been successfully used to extend the Crab Nebula differential energy
spectrum from few hundred GeV up to 80\,TeV 
\citep[][HEGRA and H.E.S.S., respectively]{aharonian:2004:hegra:crab,aharonian:2006:hess:crab}. 
However, the 
spectrum below 200 GeV has been observed only recently, 
revealing the long-anticipated IC peak in the distribution.
At low energies, space-based instruments, like \emph{Fermi}-LAT, have
improved the sensitivity in the energy range between few and
hundred GeV \citep{abdo:2010:fermi:crab}; whereas, on the other side, 
ground-based imaging atmospheric Cherenkov telescopes (IACTs) with
larger reflective surface reached lower energy
thresholds, below 100 GeV. 
The observations carried out by 
the stand-alone first MAGIC\footnote{Major Atmospheric Gamma Imaging
  Cherenkov} telescope (MAGIC-I) showed a hardening of the spectrum
below a few hundred GeV \citep{albert:2008:magic:crab}. 
However, in previous studies using MAGIC-I and \emph{Fermi}-LAT measurements,
the spectral overlap required to make a precision measurement of the
IC peak energy was not achieved.
Moreover, the quality of the available data around the IC peak 
was insufficient to rule out existing PWN models or at least distinguish between them.
The goal of this work is to use the MAGIC stereoscopic system to
measure, with high statistical precision, the Crab Nebula differential
energy spectrum down to energies of 50 GeV, and to compare this spectral
measurement with state-of-the-art PWN models.

The Crab Nebula was adopted as a standard candle in many energy
regimes, due to its high luminosity and apparent overall long-term flux stability.
It has been used to cross-calibrate X-ray and
$\gamma$-ray telescopes, to check the instrument performance over 
time, and to provide units for the emission of other astrophysical
objects. However, 
in 2010 September, both \emph{AGILE}  and \emph{Fermi}-LAT detected an 
enhancement of the $\gamma$-ray flux above 100~MeV 
\citep{tavani:2011:agile:science:crabflare, abdo:2011:fermi:science:crabflare}. 
Variability has also been measured in X rays on yearly time scale \citep{wilson-hodge:2011a}.
A search for possible flux variations in MAGIC data coinciding with the GeV flares
will be discussed in a separate paper.

\section{Observations and analysis}

MAGIC currently consists of two 17~m diameter IACTs located in the Canary Island
of La Palma (Spain) at a height of 2200~m above sea level. It is
sensitive to very-high-energy (VHE) $\gamma$ rays in the energy range
between a few tens of GeV and a few tens of TeV.
MAGIC started operations in autumn 2004 as a single telescope, MAGIC-I, 
and became a stereoscopic system five years later in 2009. During  the summers 
2011 and 2012, MAGIC underwent a major upgrade involving the readout systems 
of both telescopes and the MAGIC-I camera \citep{aleksic:2014:magic:stereo:upgrade}.
The stereoscopic observation mode led to a significant improvement in the 
performance of the instrument with an increase in sensitivity by a factor of more than two, 
while the upgrade, meant to equalize the performance of the two telescopes, 
improved the sensitivity of the instrument mainly at energies below 200\,GeV
\citep{aleksic:2014:magic:stereo:performance}.

In this work we use MAGIC stereoscopic observations of the Crab Nebula 
carried out between October 2009 and April 2011, before the above-mentioned
upgrade\footnote{Data after the upgrade are currently being studied
  and will be matter of a forthcoming publication.}. 
The instrument performance in this period, described in detail in 
\citet{aleksic:2012:magic:stereo:performance}, was sufficient to
measure a point-like source with a power-law photon index of 2.6 and
an integral flux of $9 \times 10^{-13}$ cm$^{-2}$ s$^{-1}$ above 300
GeV, at 5-$\sigma$ in 50\,h of low zenith angle observations.
The selected data set includes observations performed in \emph{wobble} mode 
\citep{fomin:1994:wobble} at zenith angles between 5$^\circ$ and 62$^\circ$. 
Data affected by hardware problems, bad atmospheric conditions,
or displaying unusual background rates were rejected to
ensure a stable performance, resulting in 69\,h of effective time.

The analysis was performed by using the tools of the standard MAGIC 
analysis software \citep{zanin:2013:magic:mars}. Each telescope records
only the events selected by the hardware stereo trigger. For every event the 
image cleaning procedure selects the pixels which have significant
signal and removes the rest. The obtained reconstructed image is then quantified with a
few parameters. 
For the analysis of the Crab Nebula data set we used \emph{sum image cleaning}, a new algorithm
which lowers the analysis energy threshold to 55\,GeV and provides a 15$\%$
improvement in sensitivity below 150\,GeV
\citep{lombardi:2011:magic:sumcleaning}.

\begin{figure*}[t!]
\centering
\includegraphics[width=4.5in]{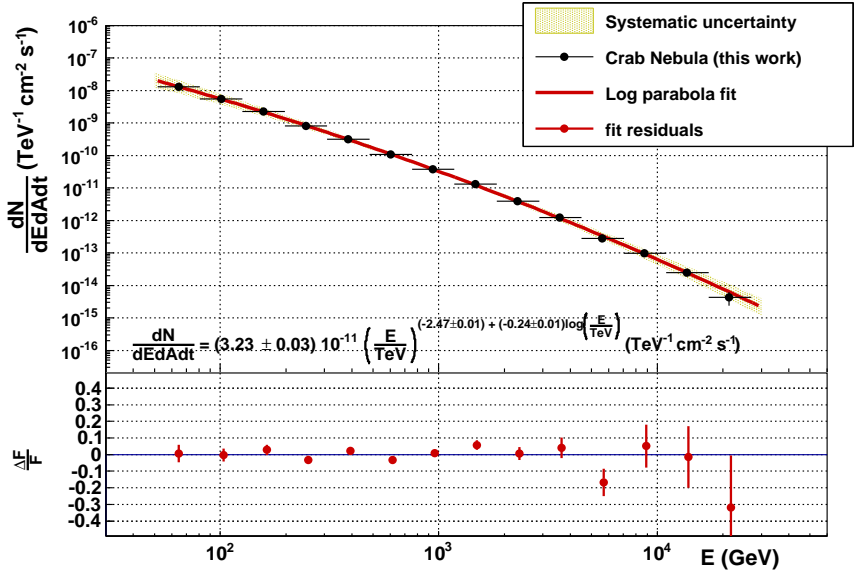}
\caption{Differential energy spectrum of the Crab Nebula obtained with
  data recorded by the MAGIC stereoscopic system.  
\label{fig:spectrum}}
\end{figure*}

After the image cleaning procedure, stereoscopic pairs of images are
combined and the shower direction is determined as the crossing point 
of the corresponding single-telescope directions. The reconstruction 
of the shower direction is later improved by applying an upgraded version of the 
\emph{disp} method \citep{zanin:2013:magic:mars}.
The background rejection relies on the definition of the multi-variable
parameter \emph{hadronness}, which is computed by means of a Random
Forest (RF) algorithm \citep{albert:2008:magic:RF}. RF uses as input a 
small set of image parameters from both telescopes, together with the 
information about the height of the shower maximum in the atmosphere 
provided by the stereoscopic reconstruction. 
The $\gamma$-ray signal is estimated through the distribution of the 
squared angular distance ($\theta^2$) between the reconstructed and 
the catalog source position. The energy of each event is estimated by 
using look-up tables created from Monte Carlo (MC) simulated 
$\gamma$-ray events. For the computation of the differential
energy spectrum, the $\gamma$-ray signal in each energy bin is
determined by selecting a soft \emph{hadronness} cut, which retains 
90$\%$ of the $\gamma$-ray events ensuring a good agreement between 
data and MC. Next, an unfolding procedure is applied to the obtained 
differential energy spectra to correct for the energy bias and the finite
energy resolution of the detector. In particular, we apply five different unfolding
methods described in \citet{albert:2007:magic:spectrumunfolding} and check the
consistency of the results. 
For the light curves, we compute integral $\gamma$-ray fluxes in a
given energy range as a function of time. No full-fledged unfolding procedure
is used here. Instead, a correction is applied to the effective area
in the selected energy range to account for the spillover of the
Monte Carlo simulated events with (true) energy outside of it, under the assumption of a
given shape of the differential energy spectrum.

Since our data set spans a large zenith angle range (5$^\circ$ to
62$^\circ$), we divide the data sample in three zenith angle
ranges\footnote{The binning in zenith angle (zd) is equidistant in
  $\cos(\mathrm{zd})$.} to better account for corresponding variations in the
image parameters: a) 5$^\circ$ to 35$^\circ$, b) 35$^\circ$ to
50$^\circ$, and c) 50$^\circ$ to 62$^\circ$.
The matrices for the background rejection obtained through the RF, as
well as the look-up tables for the energy estimation, are computed 
separately for each sub-sample. The three independent analyses are
then combined with the spectral unfolding procedure.

\section{Results}
\label{sec:SED}
\subsection{The differential energy spectrum}

The main result of this work, shown in Figure \ref{fig:spectrum}, 
  is the differential energy spectrum of the Crab Nebula obtained with
  a single instrument covering almost three decades in energy, from
  50\,GeV up to 30\,TeV, and spanning seven orders of magnitude in
  flux. It is
unfolded with Tikhonov's method \citep{tikhonov:1977}, but all the
other considered unfolding methods provide compatible results within
the statistical errors. 
The spectrum has five spectral points
per energy decade and statistical errors as low as 5$\%$ below 150\,GeV. 
Below 10\,TeV, the overall uncertainty is dominated by systematic,
rather than statistical, uncertainties. The systematic uncertainties, displayed in Figure 
\ref{fig:spectrum} as the shaded area, will be discussed in detail below.

The overall IC emission from the Crab Nebula, as well as
many other PWNe, is usually approximately described by a log-parabola
function ($dN/dE = f_{0}\,\; \left ( \frac{E}{E_{0}} \right )
^{-\alpha + \beta \log  \left ( {E}/{E_{0}} \right )} $). However,
other functional forms can provide good fits of the
measured VHE emission from the Crab Nebula over specific energy
sub-ranges. In literature, between $\sim$0.5 and 50--80\,TeV, the Crab Nebula
spectrum was described by either a power law \citep[$dN/dE = f_{0}\,\; \left (
  \frac{E}{E_{0}} \right ) ^{-\alpha} \; $]{aharonian:2004:hegra:crab}
or a power law with an exponential cut off
\citep [$dN/dE = f_{0}\,\; \left ( \frac{E}{E_{0}} \right ) ^{-\alpha} \;
\exp \left( -\frac{E}{E_{\mbox{\scriptsize c}}} \right)
$]{aharonian:2006:hess:crab}. We considered both a log-parabola and a 
power law with an exponential cut off for the analytical description
of the new spectral energy points presented in this work. The
single power-law function fails in representing them over the wide
energy range covered by the new MAGIC measurement due to an obvious
curvature in the measured spectrum.

The fits do not include systematic uncertainties, but they take into 
account the correlations between the spectral energy points. 
The power law with exponential cut off (not shown in the figure) results in a flux normalization 
$f_0 = ( 3.80\pm0.11 )$ $10^{-11}$ TeV$^{-1}$ cm$^{-2}$ s$^{-1}$, a photon index $\alpha = 2.21\pm0.02 $, 
and a cut off at $E_c = ( 6.0\pm0.6 )$\,TeV with a $\chi^2_{\mathrm{red}}$ of 35/11.
The low fit probability is mainly due to the disagreement between the
sharp cut off predicted by the fit function and the MAGIC data.
As a result the three highest flux points lie above the fit function. Excluding them and repeating the fit
we obtain a good fit quality of $\chi^2_{\mathrm{red}}$ = 8/8.
The fit to the log-parabola gives a flux normalization $f_0 = ( 3.23\pm0.03 )$ $10^{-11}$ 
TeV$^{-1}$ cm$^{-2}$ s$^{-1}$, a photon index $\alpha = 2.47\pm0.01 $, and a curvature parameter 
$\beta = -0.24\pm0.01 $.
It has a $\chi^2_{\mathrm{red}}$ of 20/11. 
The energy $E_{0}$ was fixed at 1\,TeV for both fits.
The log-parabola provides a better fit compared to the power law with
exponential cut off.
In the bottom panel of Figure~\ref{fig:spectrum}, one can see residuals between our measurements and the
best fit. 
The fit results for the power law with exponential cut off and log-parabola
are summarized in Table~\ref{tab:fit}.\\

\begin{figure*}[t!]
\centering
\includegraphics[width=6.0in]{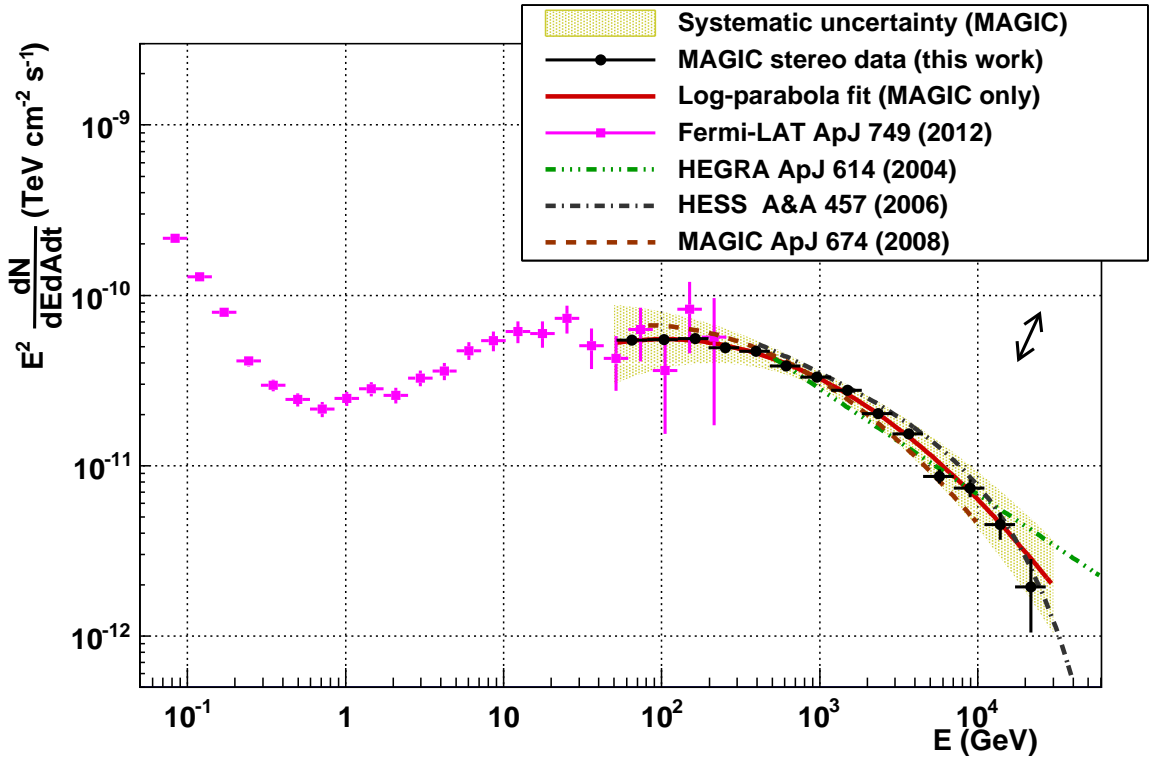}
\caption{Spectral energy distribution of the Crab Nebula from 100\,MeV
  to $\sim$30\,TeV obtained by \emph{Fermi}-LAT and MAGIC, together
  with the fit results from other $\gamma$-ray experiments. 
  The black arrow indicates the systematic
  uncertainty on the energy scale, whereas the shaded area indicates
  the systematic uncertainty on the flux normalization and the photon index.
  The solid red line is the log-parabola fit to the MAGIC data alone (the same as in Fig.\ref{fig:spectrum}).
\label{fig:SED}}
\end{figure*}

\begin{table}[htb]
\centering
\footnotesize
\begin{tabular}{lcc}
\hline
{\bf Parameter} & {\bf Power law with cutoff} & {\bf Log-parabola}\\
\hline
\hline
$f_0$ (TeV$^{-1}$ cm$^{-2}$ s$^{-1}$)   &  $( 3.80\pm0.11 )$ $10^{-11}$   & $( 3.23\pm0.03 )$ $10^{-11}$ \\
\hline
index $\alpha$ & $2.21\pm0.02 $ &  $2.47\pm0.01 $ \\
\hline
curvature $\beta$ & ---  &  $-0.24\pm0.01$ \\
\hline
cutoff $E_c$ (TeV) & $6.0\pm0.6$    & --- \\
\hline
$\chi^2_{\mathrm{red}}$  & 35/11    & 20/11 \\
\hline
\hline
\end{tabular}
\caption{Best-fit parameters to the differential photon spectrum of the Crab Nebula obtained with MAGIC
in the energy range between 50\,GeV and 30\,TeV.}
\label{tab:fit}
\end{table}

The overall systematic uncertainty affecting the measurement of the
differential energy spectrum of the Crab Nebula includes three
different classes of effects: one on the energy scale, the second in
the flux normalization and the third on the spectral shape. We
  considered all the sources of systematic uncertainty stated in Table 4
in \citet{aleksic:2012:magic:stereo:performance}, and, in addition,
the effect of the different zenith angle observations.
The uncertainty on the energy scale is 15--17$\%$,
and for the flux normalization is about 11$\%$
\citep{aleksic:2012:magic:stereo:performance}. 
The estimation of the systematic error on the spectral shape is unique
to this work since we further split the error into an uncertainty on the
photon index and one on the curvature parameter, given the assumed
log-parabola spectral shape. Both include a common uncertainty of 0.04 due to 
the non-linearity of the analog signal chain \citep{aleksic:2012:magic:stereo:performance} 
and an individual uncertainty due to the analysis methods. The latter is
evaluated as the RMS of the distributions of the $\alpha$ and the
$\beta$ parameters derived from different analyses performed with
various RFs, different image cleaning algorithms, observation zenith
angles, and efficiency of $\gamma$-ray selection cuts. This yields 
a systematic uncertainty on $\alpha$ of 0.03 and on $\beta$ of 0.05. 
The overall systematic uncertainty for both $\alpha$ and $\beta$ is
calculated by summing up in quadratures these values 
to the above-mentioned uncertainty of 0.04 for the effect of the
non-linearity, obtaining an overall of 0.05 and 0.07 for $\alpha$ and $\beta$,
respectively.

\subsection{Spectral energy distribution of the Crab Nebula}

\begin{figure}[h!]
\centering
\includegraphics[width=\linewidth]{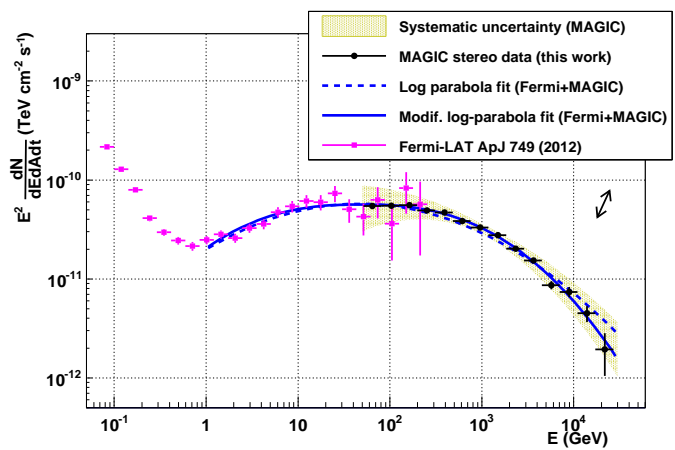}
\caption{Spectral energy distribution of the Crab Nebula obtained by
  \emph{Fermi}-LAT and MAGIC. The two lines indicate the results of
  the fits to the combination of \emph{Fermi}-LAT and MAGIC spectral
  points, see text for details.
\label{fig:SEDfit}}
\end{figure}

Figure \ref{fig:SED} shows the spectral energy distribution (SED) for
the MAGIC data (same data set as used for Figure~\ref{fig:spectrum}), and compares it to
the measurements by other IACTs (green, black and brown lines) 
as well as to the \emph{Fermi}-LAT results for the Crab Nebula (magenta squares).  
In this work we used the latest \emph{Fermi}-LAT published results
on the Crab Nebula, which include 33 months of data \citep{buehler:2012a}.
At low energies (50--200\,GeV), MAGIC data
overlaps with the \emph{Fermi}-LAT measurements, showing an
agreement, within the statistical errors, between the spectral
points of the two instruments. 
At higher energies (above 10 TeV), a disagreement between
HEGRA \citep{aharonian:2004:hegra:crab} and H.E.S.S.\ 
\citep{aharonian:2006:hess:crab} measurements has been noted (green dash-triple-dotted
and black dash-dotted lines, respectively).
This may be due to systematic uncertainties between the two instruments
or may indicate a real spectral variability of the nebula. 
The relatively large systematic uncertainty of
the MAGIC measurement and the lack of MAGIC data above 30\,TeV
do not support either hypothesis. Since the new MAGIC spectrum is
  statistically limited at these energies, we may improve the result in
the future by taking a significant amount of additional Crab Nebula
data with MAGIC.

\begin{figure*}[t!]
  \centering
  \includegraphics[width=7.0in]{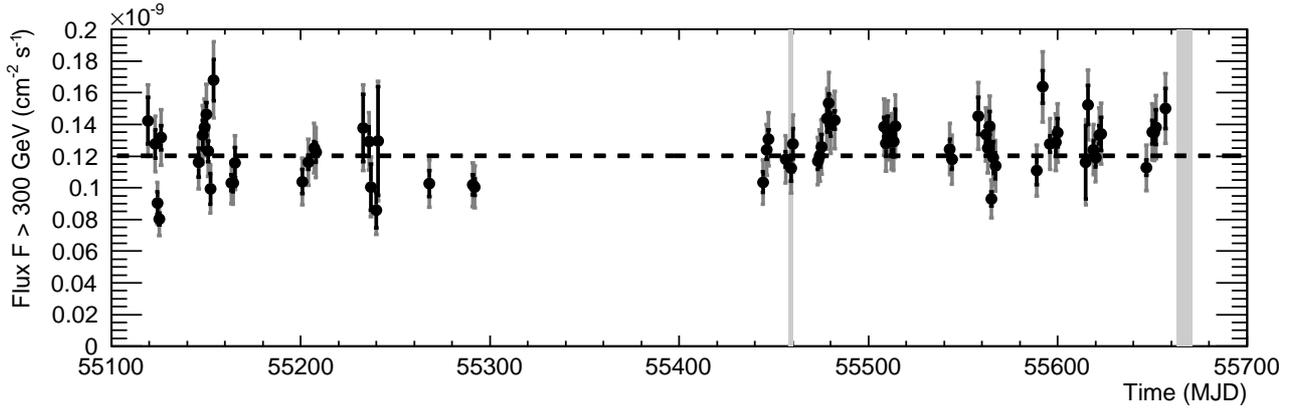}
  \caption{Daily light curve of the Crab Nebula for energies
    above 300 GeV.
The black vertical lines show statistical error bars, the grey ones the quadratic sum of statistical errors and a 12\% systematic point-wise uncertainty. 
The dashed horizontal line is the best fit value to a constant flux.
The grey areas indicate the Crab flares as reported by AGILE and \emph{Fermi}-LAT.}
  \label{fig:LC}
 \end{figure*}

To ensure independence from theoretical modeling, and assuming
that the easiest approximation for the IC contribution of the Crab
Nebula emission is a log-parabolic shape, we estimate the position of the IC peak
by a log-parabola fit to the data.
In all fits described below we take the correlations between MAGIC
spectral points into account and consider only statistical errors
unless stated otherwise.
The fit to the MAGIC data alone can locate the IC peak and doing
so results in 
(103$\pm$8)\,GeV ($\chi^{2}_{\mathrm{red}}=20/11)$, consistent with
the earlier single telescope MAGIC result \citep[(77$\pm$47)\,GeV,][]{albert:2008:magic:crab}.
A more robust fit result is obtained by considering
of MAGIC and \emph{Fermi}-LAT spectral data together since they cover both sides of the IC peak.
We, therefore, perform a joint fit to the 
MAGIC and \emph{Fermi}-LAT spectral data points
starting from 1\,GeV, corresponding to the energy of the lowest spectral
point of the \emph{Fermi}-LAT spectrum where the IC contribution
dominates over synchrotron emission. The best fit result
($\chi^2_{\mathrm{red}}=82/27)$ is shown as dashed line in Figure~\ref{fig:SEDfit}. It
results in an IC peak position at (53$\pm$3)\,GeV. 

In the following we investigate how systematic uncertainties of
the instruments may alter the fit result.
First, we include an ad-hoc point-wise flux uncertainty to MAGIC data.
  We would need an additional point-wise flux uncertainty of $>25\%$ in order
  to obtain an acceptable fit with a probability larger than 5\%. Such
  ad-hoc uncertainty exceeds both MAGIC and \emph{Fermi}-LAT
  systematic errors on the flux normalization.
Second, we allow a shift in the energy scale of the MAGIC data relative to the
  \emph{Fermi}-LAT data\footnote{We consider \emph{Fermi}-LAT to be better calibrated
  since it was absolutely calibrated with test beams at CERN before launch
\citep{altwood:2009a}, whereas there is no test beam for the IACT
technique.}, the best fit ($\chi^{2}_{\mathrm{red}}=74/26)$
locates the IC peak at (69\pmi7)\, GeV, for a +11\% shift. 
Third, we consider bracketing cases in the MAGIC systematic uncertainty
in the energy scale (15\%), in the MAGIC flux normalization (11\%), 
in the \emph{Fermi}-LAT flux normalization (5\%),
and in the \emph{Fermi}-LAT energy scale (+2\% and -5\%).
The resulting IC peak positions using any
combination of the considered uncertainties range from 40\,GeV
up to 84\,GeV. Thus, we determine the IC peak position
to ($53 \pm 3_{\mathrm{stat}} +31_{\mathrm{syst}} -13_{\mathrm{syst}}$)\,GeV
including the systematic uncertainties of the two instruments and
assuming that the peak can be described by a log-parabola.
However, none of the combinations (fits performed) resulted in an acceptable fit quality.
The highest fit probability obtained is $10^{-5}$.
We, therefore, conclude that the quality of the data presented here shows
clearly that the log-parabola cannot be used to describe the IC peak over an
energy range spanning four decades even considering systematic uncertainties of 
MAGIC and \emph{Fermi}-LAT.

To further conclude on the actual IC peak position we investigated different fit ranges 
and also looked for
a spectral model which better reproduces the new observational results.
We find that the log-parabola is a good fit ($\chi^2_\mathrm{red} = 14/13$)
if considering data in a small region around the peak only, namely between
5\,GeV and 500\,GeV: The IC peak is then at $(51 \pm 11)$\,GeV. 
To improve the likelihood of the fit in the whole IC component regime (1\,GeV -- 30\,TeV), we
considered functions with extra free parameters. The most
satisfactory fit is achieved using a modified log-parabola function:\\
 $E^2 \times dN/dE = 10^{\log f_{0}+ C\left(\log \left(\frac{E}{E_{IC}}\right)\right)^a}$.
Such a fit function, with one more
free parameter $a$ than a log-parabola discussed above, provides acceptable results with a
$\chi^{2}_{\mathrm{red}}=35/26$, locating the IC peak at
$E_{IC}=(48\pm2)$\,GeV. The resulting exponent $a=2.5\pm0.1$ produces a flatter
peak than the one obtained by the canonical quadratic
function, see both fit functions in Fig.~\ref{fig:SEDfit}. 
The other fit parameters are: $C=-10.248 \pm 0.006$
and $\log(f_{0})=-0.120 \pm 0.008$, both in units of
$[log(\mathrm{TeV/cm2/s})]$. 
Also a power law function with a sub-exponential cutoff \\
$E^2 \times dN/dE = N_{0}\,\; \left (\frac{E}{E_{0}} \right ) ^{-\alpha} \exp \left(- E/E_\mathrm{cutoff} \right)^{\beta} $
provides an acceptable fit
($\chi^{2}_{\mathrm{red}}=39/26$) with 
$N_0 = (6.8 \pm 0.6)\,\mathrm{TeV cm}^{-2} \mathrm{s}^{-1}$,
$\alpha = 1.59 \pm 0.02$, $E_\mathrm{cutoff} = (20.8 \pm 3.9)$\,GeV
and $\beta = 0.285 \pm 0.006$. The maximum in such mathematical approximation
is reached at 76\,GeV.

Even though the fit functions above provide a good fit to the joint data set
without any shift in energy scale or flux normalization, they are not
physically motivated.  We note that the fit functions and fit ranges we
exploited here yield a peak position within the systematic uncertainties of the
log-parabola fit stated above.

\subsection{The light curve}
\label{sec:LC}
In this section we present the light curve above 300 GeV from the Crab Nebula.
This is meant to check the flux stability on time scales of days.
The results are presented in Figure \ref{fig:LC}, which shows the MAGIC daily fluxes 
between October 15, 2009 and April 6, 2011, where the error bars indicate statistical (shown in black) 
and systematic errors (the combined error is shown in grey).
The average flux above 300 GeV F$_{>300\mathrm{GeV}}$ is:
\begin{equation*}
\mathrm{F}_{>300\mathrm{GeV}} = (1.20 \pm 0.08_{stat} \pm
0.17_{sys}) \times 10^{-10} \mathrm{cm^{-2} s^{-1}}
\end{equation*}
The systematic error on the integral flux is estimated to be 14\%,
excluding any possible shift in the energy scale.
The derived Crab Nebula flux is stable (fit by a constant has a probability of 15\%)
within statistical errors and a 12\% systematic point-wise uncertainty,
added in quadrature. This agrees with the systematic uncertainty expected 
for run-to-run data obtained in \citet{aleksic:2012:magic:stereo:performance}.
Note that the systematic uncertainty in
\citet{aleksic:2012:magic:stereo:performance} was computed using the
same source, the Crab Nebula. 
Thus, we cannot completely exclude the intrinsic variability at a
level below 12\%. This point-wise systematic uncertainty is
attributed mainly to the transmission of the atmosphere for the
Cherenkov light, which can change on a daily basis or even faster due
to variations in the weather conditions, and the mirror reflectivity,
which can change due to the deposition of dust.
The grey areas correspond to the Crab flares at energies above 100 MeV as reported by AGILE
and \emph{Fermi}-LAT. 
MAGIC observed the Crab Nebula simultaneously during the flare
that occurred on MJD = 55458 -- 55460\footnote{The MAGIC data are centered
around MJD = 55459.2} but 
no enhanced activity above 300 GeV was detected.

\section{Discussion}

There are two broad classes of PWN models which have been used to
  describe the observed broad band synchrotron and IC emission of PWNe:
  models which consider the MHD solution of the downstream flow 
\citep{kennel:1984a:crab,deJager:1992:crab,atoyan:1996a:crab,deJager:1996:crab,delzanna:2006,volpi:2008a,meyer:2010a},
  and models with a simplified one-zone approach either with a constant and
  isotropic magnetic field in a static setting
  \citep{hillas:1998a,aharonian:2004:hegra:crab,meyer:2010a} or
  tracing the PWN evolution
  \citep{bednarek:2003:crab,bednarek:2005a:crab,martin:2012a}.

  The broad-band SED of the Crab Nebula has been tested against models
  in the two categories:

\begin{itemize}
  \item an MHD flow model assuming a spherical symmetry as in
    \citet{kennel:1984a:crab} and presented in \citet{meyer:2010a}.
  \item a model based on the one first suggested by
    \citet{hillas:1998a} assuming a static, constant magnetic field,
    $B$ and described in \citet{meyer:2010a}.
  \item a time-dependent spherically symmetric (1D) PWN spectral model 
    presented in \citet{martin:2012a}.
\end{itemize}

\subsection{MHD flow model}

The MHD flow model is based on the analytical modelization of the
structure of the downstream pulsar wind for the simplified case of
a spherical symmetric ideal MHD flow, as in \citet{kennel:1984a:crab}. 
The solution of the MHD flow depends on the magnetization parameter
$\sigma$, the (known) spin-down power of the pulsar, and the position 
of the TS $r_{TS}$. The injection spectrum of the particles is
parameterized to be a power-law with a cutoff and is left free to fit
the observed emission spectrum. The observed emission spectrum is
calculated by self-consistently calculating the synchrotron emissivity
of the electron distribution which is carried by the flow taking into
account synchrotron and adiabatic cooling losses. The resulting photon
density is used to calculate the IC emissivity. Also the emission of
the dust is considered as additional photon field, as described in
\citet{meyer:2010a}. The resulting spectrum is compared with the
measurements and the free parameters are minimized using a $\chi^2$
cost function. For an assumed value of $\sigma=0.0045$
\citep{meyer:2010a}, $r_w=0.14$~pc, and $L_\mathrm{sd}=5\times
10^{38}$~ergs/s the minimum value of
$\chi^{2}_{\mathrm{red}}$=2.3/254, including a relative systematic
uncertainty of $7~\%$.  While the fit is reasonably good for the
synchrotron part of the spectrum, the IC part adds a $\Delta
\chi^2\approx 200$ to the fit, indicating that the spatial structure
of the magnetic field is not consistent with the data.

\subsection{Static, constant $B$-field model}

\begin{figure*}[t]
  \includegraphics[width=3.3in]{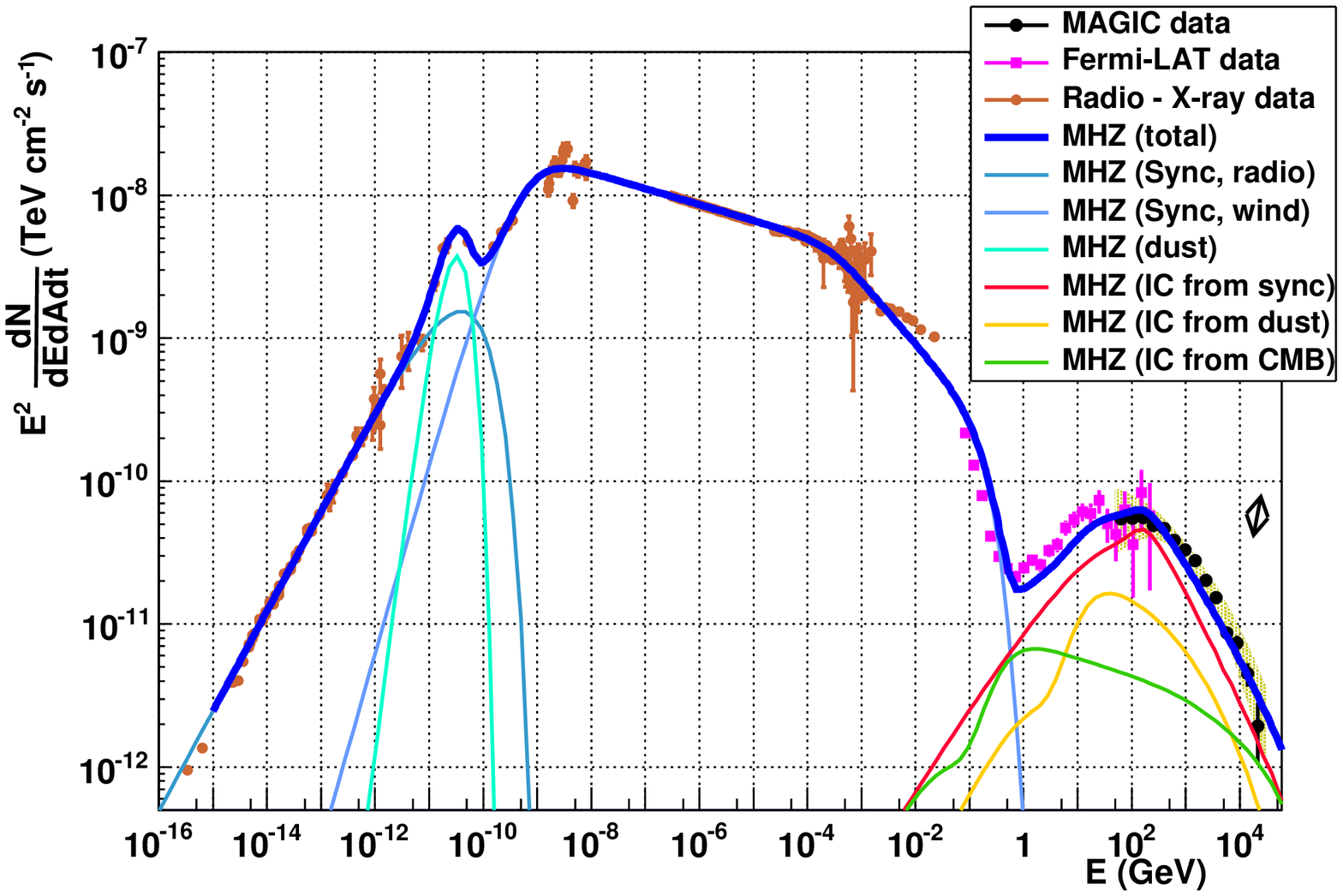}
  \includegraphics[width=3.3in]{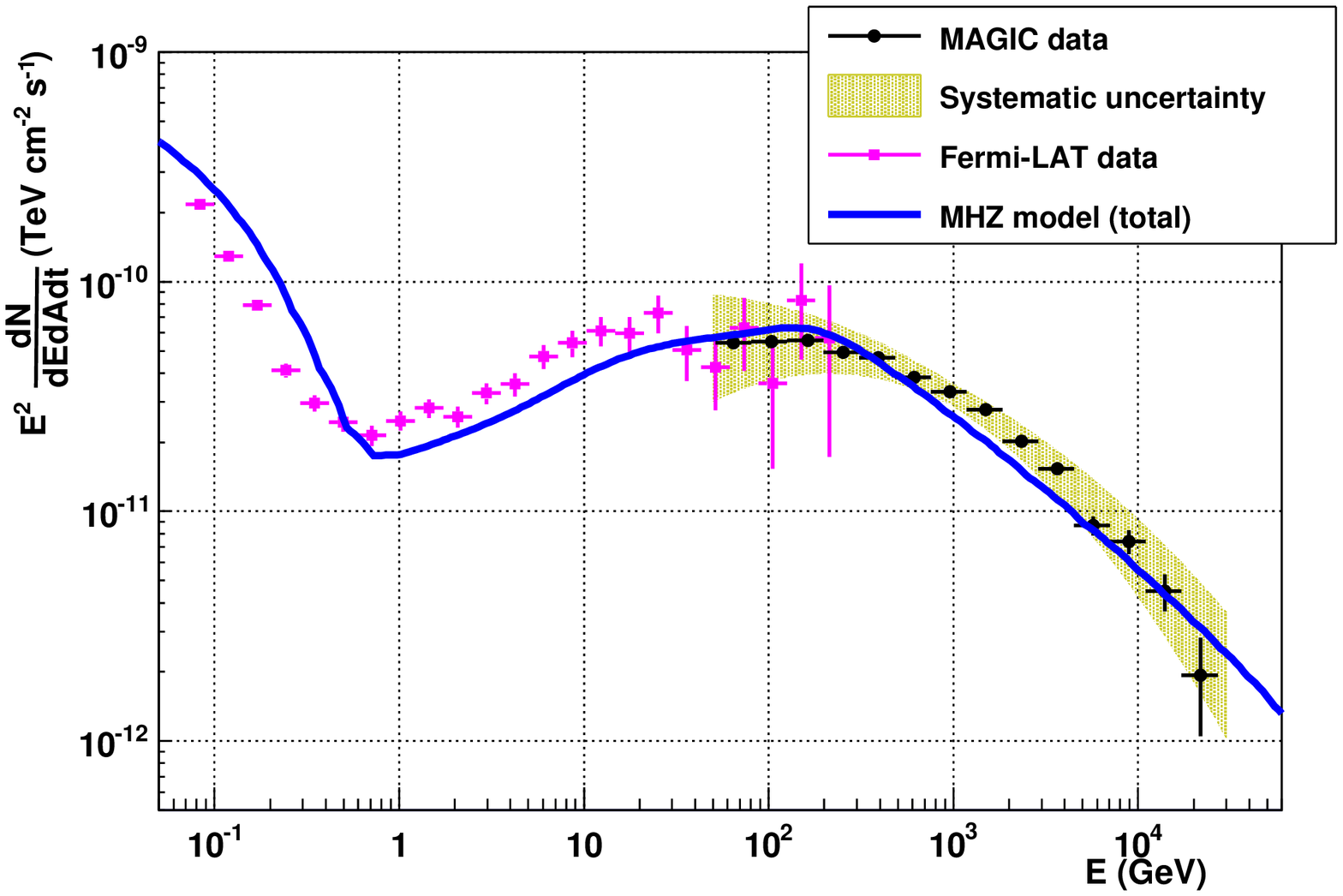}  
\caption{\emph{On the left:} The overall spectral energy distribution of the Crab Nebula
from radio to $\gamma$ rays. Lines are best fit results
based on the model of \citet{meyer:2010a} (MHZ), see text for details. The thin lines show individual components
of the photon spectrum (see the inlay), and the thick blue line identifies the overall emission.
Historical data (brown) are from \citet{meyer:2010a}, \emph{Fermi}-LAT data (pink) are from
\citet{buehler:2012a}, and the VHE data are from this work. \emph{On
  the right:} Zoom in the $\gamma$-ray regime.
  \label{fig:modelMeyer}}
 \end{figure*}

The \textit{constant $B$-field model} was introduced in
\cite{meyer:2010a} and follows the prescription put forward in
\cite{hillas:1998a} and \cite{aharonian:2004:hegra:crab}.  
The Crab Nebula is assumed to be homogeneously filled with a constant
magnetic field and two distinct electron populations: relic electrons 
(responsible for the radio synchrotron emission) and wind electrons.
The relic electron population is needed
to explain the break in the synchrotron spectrum at optical 
wavelengths \citep[see also Section\,6 in][]{meyer:2010a}. 
The relic electrons might be the result of a rapid spin-down phase
in the early stages of the evolution of the Crab Nebula
\citep{atoyan:1999:crab}. The populations can be regarded as averaged
representations of the electron distributions. The two spectra were
modeled with a simple power law and a broken power law with a
super-exponential cut off for relic and wind electrons, respectively.
For their definition we refer the reader to \citet[Eqs.\ (1) and (2) in][]{meyer:2010a}. 
The minimal gamma factor of the relic electrons
was fixed to 3.1 in the fit as it is not constrained by the observable part of the SED.
Following \citet{hillas:1998a}, the spatial distributions of both the seed
photons and pulsar wind electrons were described with Gaussian functions
in distance to the nebula's center \citep[see discussion and Eqs.\ (A.1) and
(A.2) in][]{meyer:2010a}, whereas the relic electron population is
uniformly distributed throughout the nebula.
The variances of the Gaussian distributions vary with energy, thus
accounting for the observed smaller size of the nebula at shorter
wavelengths. 
The thermal dust emission was assumed to follow a gray body spectrum.
Its extension was fixed in the fit \citep[$\theta_\mathrm{dust}
=1.3^\prime$ following][]{hillas:1998a}, while, in contrast to
\citet{meyer:2010a}, the values of remaining dust parameters were
allowed to float. 

\begin{table}
\centering
\small
\caption{Best-fit parameters for the
  constant $B$-field model. The definition of the model parameters is given in \citet{meyer:2010a}.}
\vspace{0.2cm}
\begin{tabular}{ll}
\hline
Magnitude & Crab Nebula\\
\hline
\hline
Magnetic field\\
\hline
$B$ ($\mu$G )  &   143 \\ 
\hline
Dust component\\
\hline
 $\ln(N_\mathrm{dust})$&     -29.9 \\ 
 $T_\mathrm{dust}$ (K)&      98 \\ 
$u_\mathrm{dust}$($\mathrm{eV}\,\mathrm{cm}^{-3}$ )&      1.2\\  
\hline
Radio electrons\\
\hline
 $S_r$	   &      1.6 \\
$\ln N_r$ &     119.8 \\
$\ln\gamma_r^\mathrm{min}$ &    3.1 \\ 
$\ln\gamma_r^\mathrm{max}$&     12  \\ 
\hline
Wind electrons\\
\hline
$S_w$ 	   &      3.2 \\ 
$\Delta S$ &      0.6 \\ 
$\ln N_w$ &      78.5 \\ 
$\ln\gamma_w^\mathrm{min}$&      12.9 \\
$1 / \ln\gamma_w^\mathrm{break}$    &     -19.5\\
$\ln\gamma_w^\mathrm{max}$&      22.7 \\
$\beta$		&       4 \\
\hline
\hline
\label{tab:const-B}
\end{tabular}
\end{table}

The electron spectra were calculated using the same synchrotron data
as in \citet{meyer:2010a} except for the new \emph{Fermi}-LAT data
\citep{buehler:2012a}.
For a given magnetic field strength, the parameters of the electron
spectra were
derived from the fit to the synchrotron data between
$4\cdot10^{-6}\,\mathrm{eV} \leqslant \nu \leqslant
0.4\,\mathrm{GeV}$, using a $\chi^2$ minimization  implemented with
the interface of \texttt{MINUIT} \citep{Minuit}. 
Subsequently, the magnetic field and the parameters describing
the thermal dust emission were varied until the IC part of the SED 
($\mathrm{E} > 0.4\,\mathrm{GeV}$) presented in this work is
reproduced best. The full Klein-Nishina cross section is used to
calculate the IC emission including synchrotron and thermal dust
emission, and the cosmic microwave background (CMB). 

Allowing for a point-wise systematic uncertainty of 8$\%$ of the flux 
\citep[added in quadrature,][]{meyer:2010a}, the synchrotron emission
is accurately reproduced with $\chi^{2}_{\mathrm{red}}=249/217=1.15$
(Figure~\ref{fig:modelMeyer}). Above $0.4\,\mathrm{GeV}$, the data is
poorly described and the fit only converges if an ad-hoc
(unrealistically large) systematic uncertainty of 17\,\%  is assumed, resulting
in $\chi^2_{\mathrm{red}} = 48.8 / 31 = 1.57$. 

The final best-fit parameters are given in Table~\ref{tab:const-B}.
Due to the small fit probability and the dependence of the fit
  errors on the additional ad-hoc systematic uncertainty
added to the flux points, we neglect these uncertainties. 
When comparing the result of \citet{meyer:2010a}
with the one presented here, $B=143\,\mu$G, we note that a higher value of the B-field is
preferred compared to the 2010 paper in order to reproduce the MAGIC data
around the IC peak. The higher quality (i.e.\ smaller error bars) of the
\emph{Fermi}-LAT data together with the MAGIC data shows a rather flat peak now,
which cannot be reproduced in the model.  If we would repeat the exact procedure
from the 2010 paper and only use the updated \emph{Fermi}-LAT data, we would
find a lower B-field and the model would undershoot the MAGIC data at almost
all energies.  We, therefore, conclude that the constant $B$-field model cannot
reproduce the flat peak of the IC SED.  For energies above the peak, the
predicted spectrum is too soft with too little curvature as compared to the
new MAGIC data.

\subsection{Time-dependent model}
\begin{table}
\centering
\small
\caption{Fit parameters for the time-dependent model obtained
  with the new data points given by {\it MAGIC}. The definition of the parameters
  can be found in \citet{martin:2012a}. 
  }
\vspace{0.2cm}
\begin{tabular}{ll}
\hline
Magnitude  & Crab Nebula\\
\hline
\hline
Pulsar magnitudes\\
\hline
$P$ (ms) & 33.40\\
$\dot{P}$ (s s$^{-1}$) & 4.21 $\times 10^{-13}$\\
$\tau_c$ (yr) & 1260\\
$t_{age}$ (yr) & 960\\
$L(t_{age})$ (erg s$^{-1}$) & 4.3 $\times 10^{38}$\\
$L_0$ (erg s$^{-1}$) & 3.0 $\times 10^{39}$\\
$n$ & 2.509\\
$\tau_0$ (yr) & 730\\
$d$ (kpc) & 2\\
$M_{ej}$ (M$_{\odot}$) & 8.5\\
$R_{PWN}$ (pc) & 2.2\\
\hline
Magnetic field\\
\hline
$B(t_{age}) ({\mu}G)$ & 80\\
$\eta$ & 0.025 \\
\hline
Wind electrons\\
\hline
$\gamma_{max} (t_{\mathrm{age}})$ & 8.3 $\times 10^{9}$\\
$\gamma_b$ & 1 $\times 10^{6}$\\
$\alpha_l$ & 1.6\\
$\alpha_h$ & 2.5\\
$\epsilon$ & 0.25\\
$R_{syn}/R_{PWN}$ & 1 \\
\hline
Target photon fields and environment density\\
\hline
$T_{FIR}$ (K) & 70 \\
$w_{FIR}$ (eV cm$^{-3}$) & 0.1\\
$T_{NIR}$ (K) & 5000\\
$w_{NIR}$ (eV cm$^{-3}$) & 0.3\\
$n_H$ (cm$^{-3}$) & 1\\
$T_{CMB}$ (K) & 2.73 \\
$w_{CMB}$ (eV cm$^{-3}$) & 0.25\\
\hline
\hline
\label{para}
\end{tabular}
\end{table}

\begin{figure*}[th]
  \centering
  \includegraphics[width=3.3in]{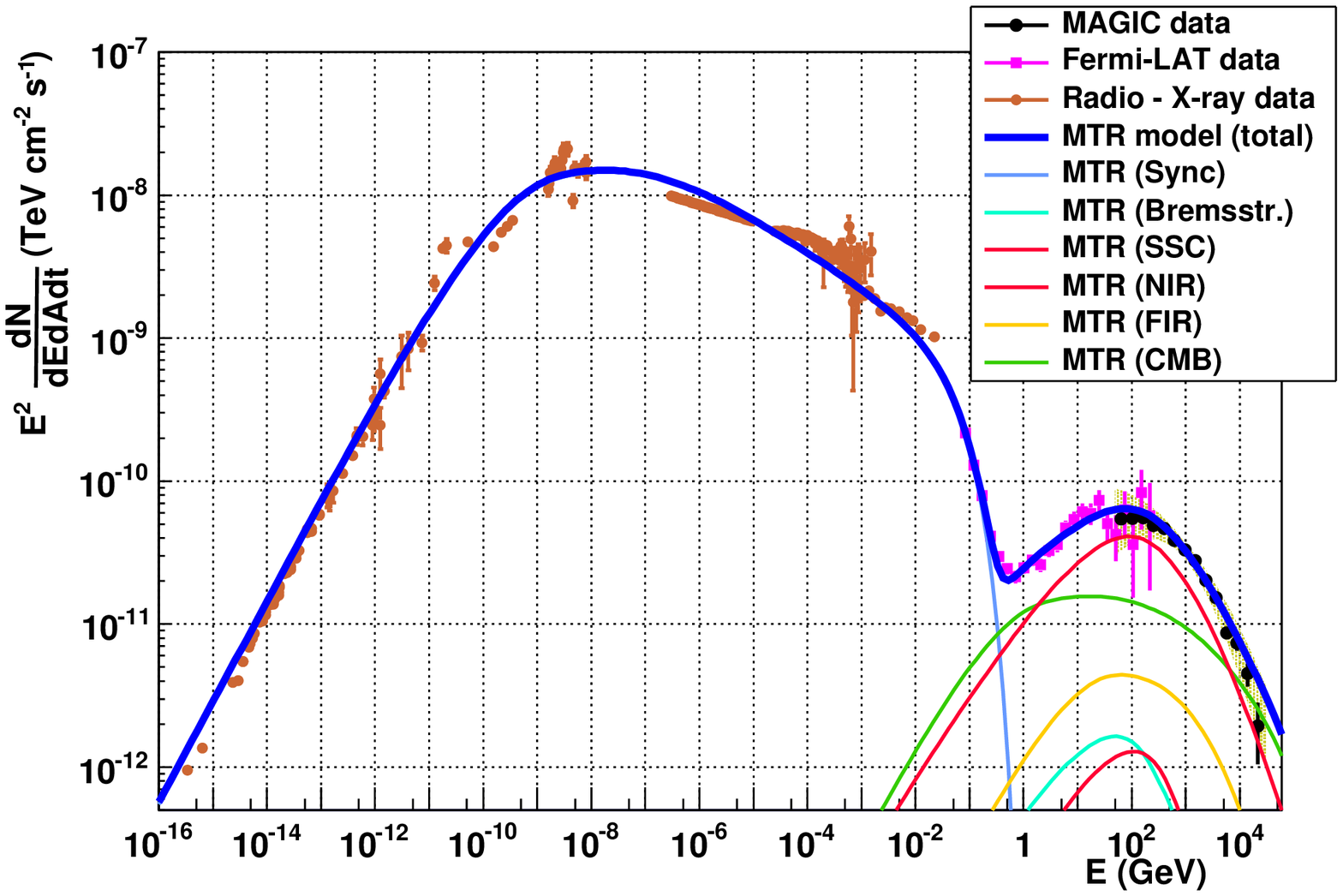}
  \includegraphics[width=3.3in]{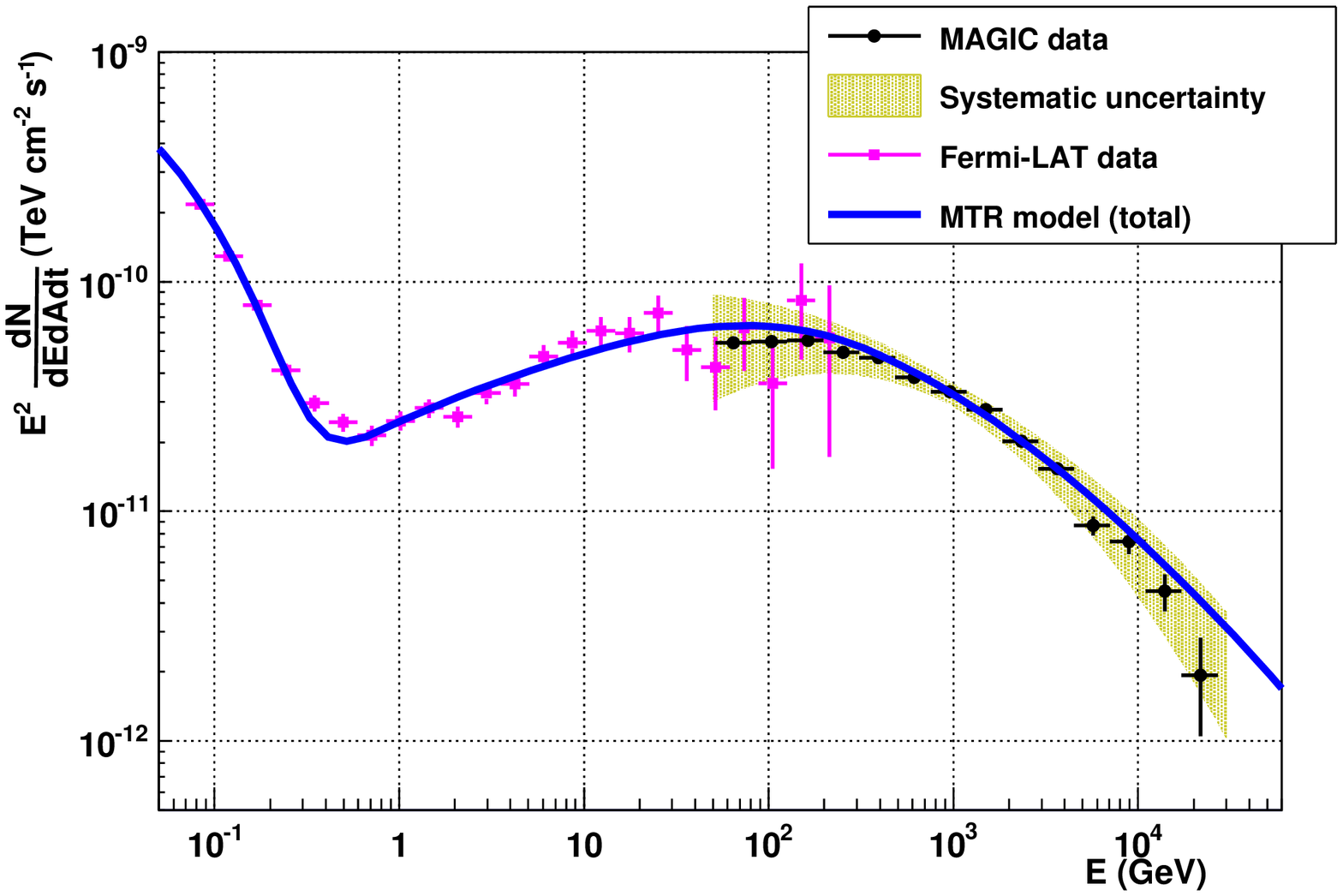}  
  \caption{\emph{On the left:} The overall spectral energy distribution of the Crab Nebula
from radio to $\gamma$ rays. Lines are best fit results
based on \citet{martin:2012a} (MTR), see text for details. The thin lines show individual components
of the photon spectrum (see the inlay), and the thick blue line identifies the overall emission.
Historical data (brown) are from \citet{meyer:2010a}, \emph{Fermi}-LAT data (pink) are from
\citet{buehler:2012a}, and the VHE data are from this work. \emph{On
  the right:} Zoom in the $\gamma$-ray regime.
  \label{fig:modelTorres}}
 \end{figure*}

The time-dependent, leptonic spectral model for an isolated PWN
\citep{martin:2012a,torres:2013a,torres:2013b} was also considered.
Such model solves the diffusion-loss equation numerically devoid of
any approximation, considering synchrotron, IC and Bremsstrahlung
energy losses. For the IC losses, the Klein-Nishina cross section
is used. Escaping particles due to Bohm diffusion are also taken
into account. The injection spectrum of the wind electrons is a 
broken power law normalized using the spin-down power of the
pulsar and the magnetic fraction\footnote{The magnetic fraction is the
  percentage of the spin down that goes into the magnetic field.}.
The 1D uniform magnetic field is evolved by solving the magnetic field
energy conservation, including its work on the environment
\citep{torres:2013b}. Considering the young age of the remnant, the
nebula was treated as freely expanding.
The magnetic fraction of the nebula ($\eta$) was assumed constant
along the evolution, and it was used to define the time-dependent
magnetic field.
The model here is essentially the same as the one shown in \citet{torres:2013a} 
except for the incorporation of a more precise dynamical evolution to
fix the nebula radius taking into account the variation of the
spin-down power in time. In particular, the evolution of the radius of
the nebula was calculated solving numerically Eq.\,(25) in
\citet{vanderSwaluw:2001a}. All other time dependent parameters 
were left free to evolve with the PWN.
The resulting electron population was used to compute the synchrotron,
IC from CMB, far infrared (FIR), and near infrared (NIR) photon fields, as well
as the synchrotron self-Compton (SSC) and bremsstrahlung spectra.

The results obtained by our qualitative fit are shown in
Figure~\ref{fig:modelTorres}, whereas the parameter values are listed
in Table~\ref{para}. The free parameters of the fit relate to 
the definition of the environment, of the wind electron
spectrum, and the magnetization. For the former, they are essentially
those describing the target photon fields with which the electrons in
the nebula interact. The parameters of the wind spectrum are those
contained in the broken power law assumed to describe the electrons.
The other parameters are fixed or strongly constrained.
Since the fit is qualitative (we are aware that by having many
simplifications the model can only be considered as qualitative
description of the nebula), we do not provide uncertainties on the fit
parameters.
We find that a low magnetic fraction of the nebula (of only a few percent)
with a magnetic field of approx. 80\,$\mu$G provides a good fit to the
nebula measurements at the current age. Such magnetic field strength is also
motivated from morphological MHD studies \citep{volpi:2008a}. 

We note some caveats regarding this model. It includes no structural information: 
the size of the synchrotron sphere is taken as the size of the nebula itself, 
at all frequencies as in, e.g., \citet{bucciantini:2011a} or in \citet{tanaka:2010a}. 
This is not the case for Crab though: the size of the nebula decreases
towards the optical frequencies, being always smaller than the one
obtained from the use of a dynamical free expansion solution. For
instance, \citet{hillas:1998a} use a radius of approximately 0.4\,pc up
to 0.02\,eV, and slightly smaller for larger energies. 
If this energy-dependent size of the synchrotron nebula is adopted 
(one-zone spheres of different sizes at different frequencies), 
the SSC emission would be overproduced. 
A full description of such a rich data set requires a more
detailed model that, in addition to being time dependent, treats 
the morphology at different frequencies using a multi-zone, multi-dimensional approach.

\section{Conclusions}

We presented a long term data set of the Crab Nebula taken with the MAGIC telescopes
between October 2009 and April 2011. We derived the differential
energy spectrum of the Crab Nebula with one single instrument,
covering almost three decades in energy, from 50\,GeV up to 30\,TeV.
The energy spectrum in this range is clearly curved
and matches well both with the \emph{Fermi}-LAT spectrum at lower energies
and with the previous Crab Nebula measurements by Whipple, HEGRA,
H.E.S.S. and early MAGIC-I data. 
The resulting IC peak is broad and rather flat in the energy range from 10\,GeV to 200\,GeV.
When considering the joint MAGIC--\emph{Fermi}-LAT fit, the function
which best describes this emission component is a modified
log-parabola (with a 2.5 exponent). Thanks to the large lever arm of the fit 
we determined the most precise IC peak position at energy
($(53\pm3_{\mathrm{stat}}+31_{\mathrm{syst}}-13_{\mathrm{syst}})$)\,GeV.
The MAGIC spectrum extends up to 30\,TeV but we cannot distinguish
between a power law tail extending up to 80\,TeV
\citep[HEGRA,][]{aharonian:2004:hegra:crab} and a spectral cutoff at
around ~14\,TeV
\citep[H.E.S.S.,][]{aharonian:2006:hess:crab}. Irrespective off
spectral variability or any other sources for this discrepancy, the
uncertainties do not permit a resolution of this issue.
We also show that the light curve of the Crab Nebula above 300\,GeV
is stable within the statistical and systematic uncertainties on the daily basis
($\sim12\%$) during the considered period. Flux stability on longer
time scales, as well as data taken simultaneously with the Crab flares
will be discussed elsewhere.

The statistical precision of the MAGIC data set, spanning for the
first time from 50\,GeV to 30\,TeV, allows for a detailed test of the
two state-of-the-art Crab Nebula models. The conclusion, based on
  earlier data, that simple models can account for the observed
  spectral shape has to be revisited in the light of the new MAGIC
  results.
The MHD flow model
\citep{meyer:2010a} assuming a spherically symmetry fails to
reproduce the IC observations, suggesting that such a simplified
structure of the magnetic field is not realistic.  The constant
B-field model \citep{meyer:2010a} leads to a rather poor fit to the
new VHE measurements, failing to reproduce the breadth of the observed
IC peak. Most probably this implies that the assumption of the
homogeneity of the magnetic field inside the nebula is incorrect. 
The time dependent 1D model by \citet{martin:2012a}
can satisfactorily reproduce the VHE data up to few TeV under
the assumptions of a low magnetic field of less than hundred $\mu$G.
It fails, however, to provide a good fit of the new spectral data
if the observed morphology of the nebula \citep[smaller size at
shorter wavelengths, as in][]{hillas:1998a} is adopted. 
Therefore, we conclude that more theoretical work on the Crab Nebula
modeling must be done to simultaneously fit the observed morphology
and the spectral energy distribution.
The broad-band IC spectrum is in principle sensitive to
the spatial structure of the magnetic field and hence can be used for
future models.

\section*{Acknowledgements}
We would like to thank the Instituto de Astrof\'{\i}sica de Canarias for the
excellent working conditions at the Observatorio del Roque de los Muchachos in
La Palma. The financial support of the German BMBF and MPG, the Italian INFN
and INAF,  the Swiss National Fund SNF, the ERDF under the Spanish MINECO, and
the Japanese JSPS and MEXT is gratefully acknowledged. This work was also
supported by the Centro de Excelencia Severo Ochoa SEV-2012-0234, CPAN
CSD2007-00042, and MultiDark CSD2009-00064 projects of the Spanish
Consolider-Ingenio 2010 programme, by grant 268740 of the Academy of Finland,
by the Croatian Science Foundation (HrZZ) Project 09/176 and the University of
Rijeka Project 13.12.1.3.02, by the DFG Collaborative Research Centers
SFB823/C4 and SFB876/C3, and by the Polish MNiSzW grant
745/N-HESS-MAGIC/2010/0.  We thank the two anonymous referees for thorough
reading and helpful comments on the manuscript.

\bibliographystyle{apj}
\bibliography{crabnebula.bbl}

\end{document}